\documentstyle[aps,eqsecnum]{revtex}
\begin{document}
\twocolumn[\hsize\textwidth\columnwidth\hsize\csname
@twocolumnfalse\endcsname
\title{%
Classical black hole evaporation in Randall-Sundrum infinite braneworld
}
\author{%
Takahiro Tanaka
}
\address{Yukawa Institute for Theoretical Physics, 
 Kyoto University, Kyoto 606-8502, Japan\\
{\rm E-mail: tanaka@yukawa.kyoto-u.ac.jp}}
\maketitle

\thispagestyle{empty}
\begin{abstract}
After the gravity induced on the brane in the Randall-Sundrum (RS) 
infinite braneworld is briefly reviewed, we discuss 
the possibility that black holes evaporate as a 
result of classical evolution
in this model based on the AdS/CFT correspondence. 
If this possibility is really the case, the existence of 
long-lived solar mass black holes will give the strongest constraint 
on the bulk curvature radius. At the same time, we can 
propose a new method to simulate  
the evaporation of a 4D black hole due to the Hawking radiation 
as a 5D process.  
\end{abstract}
\vspace*{10mm}]

\section{Introduction}
Current candidates for the fundamental theory of particle physics
such as string theory or M-theory are all defined as a 
theory in higher dimension. 
To obtain an appropriate 4-dimensional 
effective theory starting with such a theory, 
a certain dimensional reduction is necessary. 
One well-known scheme of dimensional 
reduction is the Kaluza-Klein compactification. 
In this scheme, the size of the extra dimension is
supposed to be very small so as not to excite the modes 
which have momentum in the direction of the extra dimension. 
This scheme seems to work well as a mechanism to shield 
the effect of extra dimensions. 
This Kaluza-Klein scheme, however, is not a unique 
possible scheme for dimensional reduction. 
Recently, the braneworld scenario has been 
attracting a lot of attention as an alternative 
possibility\cite{Horava:1996qa,Arkani-Hamed,Randall:1999ee,Randall:1999vf}. 
The essential feature of the braneworld scenario distinct from the 
ordinary Kaluza-Klein compactification is that 
the matter fields of the standard model are supposed to be localized on 
the brane, while the graviton can 
propagate in a higher dimensional spacetime which  
we call ``bulk''. 
Owing to the assumption that the 
ordinary matter fields are localized on the brane 
the braneworld models can be consistent with the 
particle physics experiments even if the length scale of 
the extra dimension is not extremely small. 
Then, the gravity is possibly altered at a 
rather macroscopic length scale, while the experimental constraint 
on the deviation of the gravitational force from the Newton's law 
obtained so far is not severe below sub $mm$ scale. 
Hence, a characteristic length scale can be as large as 
sub $mm$ scale in the braneworld scenario. 
Therefore future experiments may detect 
an evidence of the existence of extra dimensions. 

In the course of studies on braneworld, a new scenario 
was proposed by Randall and Sundrum 
(RS)\cite{Randall:1999ee,Randall:1999vf}. 
In their models, 
the gravity is effectively localized due to the warped 
compactification even though the 
extension of the extra dimension is infinite.  
So far any results significantly distinguishable from the 
4D general relativity have not been reported. 

In this paper, we first review the current status of the 
studies on the gravity in this model. 
In \S 2 we explain the setup of the RS model with 
infinite extra dimension. In \S 3 we review the geometrical 
approach to the gravity in this model to find 
the limitation of the analysis without solving the 
5D equations of motion. In \S 4 and \S 5 
we summarize the results for linear perturbations 
and for non-linear perturbations of this model, 
respectively. Even in the non-linear regime, 
4D general relativity seems to be recovered. 
However, black hole solutions as a symbol of strong gravity 
have not been found yet. 
In \S6 we discuss the possibility that there is 
no static black hole solution in this model, applying 
the argument of the AdS/CFT correspondence. 
In \S 7 we discuss the importance of the conjectured possibility.  
If the conjecture is true,  the existence of 
long-lived solar mass black holes will give the strongest constraint 
on the bulk curvature radius. At the same time, we can 
propose a new method to simulate  
the evaporation of a 4D black hole due to the Hawking radiation 
as a 5D process.

\section{Warped extra dimension}
In the model proposed by Randall and Sundrum\cite{Randall:1999vf}, 
5D Einstein gravity with negative cosmological 
constant $\Lambda$ is assumed. 
The ordinary matter fields are 
confined on a 4-dimensional object called ``brane''. This brane 
has positive tension $\sigma$, and the spacetime has 
reflection symmetry ($Z_2$-symmetry) at the position 
of the brane, $y=y_b$. Here $y$ is the coordinate normal to the brane.  
5D Einstein equations are 
\begin{equation}
 ^{(5)}G_{ab}=-\Lambda g_{ab}+8\pi G_5 
  S_{ab} \delta(y-y_b), 
\label{5Deq}
\end{equation}
with
\begin{equation}
  S_{ab}=-\sigma\gamma_{ab}+T_{ab}, 
\end{equation}
where $T_{ab}$ is the energy momentum tensor of the 
matter field localized on the brane,$\gamma_{ab}$ is 
the 4D metric induced on the brane, and $G_5$ is the 5D Newton's constant.    

One solution of (\ref{5Deq}) 
is 5D anti-de Sitter (AdS) space 
\begin{equation}
 ds^2= dy^2 + e^{-2|y|/\ell} (-dt^2 + d{\bf x}^2), 
\end{equation}
with a single positive tension brane located at $y=0$. 
Here, $\ell$ is the curvature radius of 5-dimensional AdS space. 
The five dimensional cosmological constant and the brane tension 
are set to $\Lambda=-6/\ell^2$ and $\sigma=3/(4\pi G_5 \ell)$.  
It is convenient to introduce conformal coordinates 
defined by $z=\mbox{sign}(y)\ell(e^{|y|/\ell}-1)$. 
In these coordinates 
the metric is expressed as 
\begin{equation}
 ds^2= {\ell\over (|z|+\ell)^2}\left(
 dz^2 +\eta_{\mu\nu}dx^\mu dx^\nu\right), 
\label{confmetric}
\end{equation}
where $\eta_{\mu\nu}$ is the 4D Minkowski metric.
The outstanding feature of this model is that the 
4D general relativity is seemingly reproduced as 
an effective theory on the brane in spite of the fact that the extension 
in the direction of the extra dimension is infinite. 

\section{geometrical approach}

A quick way 
to see why the 4D general relativity is expected to 
be realized on the brane will be the 
geometrical approach introduced by Shiromizu, 
Maeda and Sasaki\cite{Shiromizu:2000wj}.  
We use the 4+1 decomposition of the 
5D Einstein tensor. The components 
parallel to the brane are decomposed by the 
Gauss equation as  
\begin{eqnarray}
 ^{(4)}G_{\mu\nu}&=&^{(5)}G_{\mu\nu} + ^{(5)}R_{yy}\gamma_{\mu\nu} 
    +K K_{\mu\nu}-K_{\mu}{}^{\rho}K_{\rho\nu}
\cr &&
    -{1\over 2}\gamma_{\mu\nu}\left(K^2-K^{\alpha\beta}K_{\alpha\beta}\right)
    -^{(5)} R_{y\mu y\nu},  
\end{eqnarray}
where $K_{\mu\nu}$ is the extrinsic curvature tensor of the 
$y=$constant hypersurfaces. 
Taking account of $Z_2$ symmetry at $y=0$, 
the Israel's junction condition gives 
\begin{equation}
  K_{\mu\nu}(y=+\epsilon)=-4\pi G_5\left(S_{\mu\nu}
      -{1\over 3}\gamma_{\mu\nu} S\right).  
\end{equation}
Substituting the above two equations into Eq.~(\ref{5Deq}), we obtain 
\begin{eqnarray}
  ^{(4)}G_{\mu\nu}=8\pi G_4 T_{\mu\nu}
               +(8\pi G_5)^2 \pi_{\mu\nu}-E_{\mu\nu}, 
\label{4DEeq}
\end{eqnarray}
where the 4D effective Newton's constant 
is given by 
\begin{eqnarray}
 G_4= {4\pi G_5^2 \sigma\over 3}={G_5\over \ell},
\end{eqnarray}
and $E_{\mu\nu}$ is a projected Weyl tensor defined by 
\begin{equation}
 E_{\mu\nu}=^{(5)}C_{y\mu y\nu}.  
\end{equation}
$\pi_{\mu\nu}$ is a tensor quadratic in $T_{\mu\nu}$ whose 
explicit form is given in Ref.~\cite{Shiromizu:2000wj}. 
By construction $E_{\mu\nu}$ is traceless. 
From the 5D conservation law for the localized matter fields, 
the 4D effective conservation law for the matter fields 
$
 ^{(4)}D_\nu\, T_\mu{}^\nu=0
$
follows, which also implies 
\begin{equation}
 ^{(4)}D_\nu \,E_\mu{}^\nu=(8\pi G_5)^2\,{}^{(4)}\!D_\nu \pi_\mu{}^\nu,  
\label{Econs}
\end{equation}
owing to the Bianchi identity. 

If we can neglect the last two terms in Eq.~(\ref{4DEeq}), 
the dynamics of 4D general relativity is recovered. 
The order of magnitude of the 
$\pi_{\mu\nu}$ term is easily evaluated to be  
smaller by the factor of 
$T_{\mu\nu}/\sigma$ than the first term, $8\pi G_4 T_{\mu\nu}$. 
Therefore we can neglect 
the $\pi_{\mu\nu}$ term at a low energy. 
On the other hand, the equation (\ref{Econs}) is not sufficient 
to determine the evolution of $E_{\mu\nu}$ in general. 
The equations that determine 
$E_{\mu\nu}$ are essentially 5 dimensional and are  
not obtained in the form of a 4D effective theory. 

\section{linear perturbation}
As we have reviewed in the previous section, 
we need to solve a 5D equation in order to 
fully determine the evolution of the metric induced on the brane. 
Since solving 5D equation in general is not easy, 
we consider linear perturbations of the RS model. 
To discuss metric perturbations in the bulk, 
the RS gauge is convenient. 
In this gauge, $y$-components of metric perturbations 
are set to be zero; $h_{y a}=0$, and also $h_{\mu\nu}$ satisfies 
the transverse and traceless conditions;
$h^{\nu}{}_{\mu,\nu}=h^\nu{}_\nu=0$.  
The homogeneous equations for bulk metric perturbations 
become
\begin{eqnarray}
&&\left[-\partial_z^2+ V(z) \right]\psi_{\mu\nu}=\eta^{\rho\sigma}
   \partial_\rho \partial_\sigma \psi_{\mu\nu}, 
\label{psieq}
\end{eqnarray}
where $\psi_{\mu\nu}=\sqrt{|z|+\ell}\, h_{\mu\nu}$ and 
\begin{equation}
 V(z)={15\over 4 (|z|+\ell)^2}-3\ell^{-1}\delta(z). 
\end{equation} 
The solution of Eq.~(\ref{psieq}) can be found 
in the form of $\psi_{\mu\nu}\propto u_m(z)e^{ik_{\mu} x^{\mu}}$. 
The separation constant $m^2=-k_{\mu}k^{\mu}$ can be understood 
as the mass of the effective 4D field which corresponds
to the mode $u_m(z)$. The equation that $u_m(z)$ satisfies is 
$
 \left[-\partial_z^2+ V(z) \right] u_m(z)=m^2 u_m(z).  
$
The solution of this equation with the $Z_2$-symmetry is 
$
u_m (z) = N_m \sqrt{|z|+\ell}(J_1(m\ell) 
     Y_2\left(m(|z|+\ell)\right)-Y_1(m\ell) J_2\left(m(|z|+\ell)\right)). 
$
The normalization $N_m$ determined so as to satisfy 
$
 2\int_{\ell}^{\infty} u_m(z) u_{m'}(z) dz = \delta(m-m')
$
is given by 
$
 N_m=\sqrt{m/ 2}\left/\sqrt{J_1(m\ell)^2+Y_1(m\ell)^2}\right.. 
$

The basic feature of these wave 
functions can be understood without resorting to 
the explicit form of the solution. 
Note that the 5D metric given in 
Eq.~(\ref{confmetric}) satisfies 5D Einstein equations  
even if we replace the Minkowski metric $\eta_{\mu\nu}dx^{\mu}dx^{\nu}$ 
in it with any vacuum solution of 4D Einstein equations.  
Corresponding to this type of solutions, 
there is a discrete mass spectrum at $m=0$ with the wave function 
$h_{\mu\nu}\propto 1/z^2$.  
We call it zero mode. This zero mode wave function is nodeless. 
Hence, there is no bound state for $m^2<0$. 
On the other hand, the potential $V(z)$ 
goes to zero at $|z|\to \infty$. Hence, the mass spectrum 
is continuous for $m^2>0$. 
The potential $V(z)$ has 
barrier near the brane with the height of $O(\ell^{-2})$. 
For the modes with $0<m\alt \ell^{-1}$, the wave function 
is suppressed near the brane due to this potential barrier. 
For the modes with $m\agt \ell^{-1}$, their excitation
is kinematically suppressed. 
Therefore the zero mode is the only active degrees of freedom, 
and it is a massless and spin-2 field in the language of 
the 4D effective theory. 
Hence, 4D general relativity is expected to be recovered at 
least at the linear level.

Linear metric perturbations induced on the brane were first 
explicitly evaluated in Ref.\cite{GarTan}. 
The result is summarized as 
\begin{eqnarray}
 h_{\mu\nu}&=& -16\pi G_5\int d^4x' G(x,x') 
   \left(T_{\mu\nu}-{1\over 3}\gamma_{\mu\nu} T\right)
\cr &&
   +{8\pi G_5\ell^{-1}\over 3}\gamma_{\mu\nu}{}^{(4)}\Box^{-1} T, 
\label{hmunu}
\end{eqnarray}
where 
\begin{eqnarray}
&& G(x,x') = -\int {d^4 k\over (2\pi)^4} e^{ik_{\mu}
    \left(x^{\mu}-x'{}^{\mu}\right)}
\cr  &&\quad \times
\left[{z^{-2}z'{}^{-2}\ell^{-1}\over k^2-(\omega+i\epsilon)^2}
    +\int_0^{\infty} \! dm
     {u_m(y)u_m(y')\over m^2+k^2-(\omega+i\epsilon)^2}\right],
\hspace{-9pt}\cr 
&&
\label{Green}
\end{eqnarray}
is the 5D scalar Green function. 
If we assume static and spherically symmetric configuration 
for the matter source localized on the brane,  
the gravitational field outside the matter distribution 
is evaluated as\cite{GarTan,Giddings:2000}
\begin{eqnarray}
&&  h_{00}\approx {2G_4 M\over r}
              \left(1+{2\ell^2\over 3r^2}\right),\cr 
&&  h_{ij}\approx {2G_4 M\over r}
              \left(1+{\ell^2\over 3r^2}\right).
\label{asymmetric}
\end{eqnarray}
The correction to 4D general relativity 
is suppressed by the ratio between the 
5D curvature scale $\ell$ and the distance from the center of the star $r$. 
The correction to the gravitational field inside the star  
also stays small by the factor of $O(\ell^2/r_{\star}^2)$, 
where $r_{\star}$ is the typical size of the star. 
If we neglect the contribution due to massive modes ($m^2>0$) in 
the Green function (\ref{Green}), Eq.~(\ref{hmunu}) 
exactly reduces to the results for the 
linearized 4D general relativity.   

\section{non-linear perturbation}
At the linear level, perturbations of the RS
model can be expressed as a 4D effective theory with an 
infinite tower of massive gravitons\footnote{There is 
an alternative way of describing the model as a 4D higher 
derivative theory\cite{ChaHaw}.}.
However, we will notice that the asymptotic behavior of the 
wave function at large $z$ 
is not very regular. The zero mode wave function behaves as 
\begin{equation}
 h_{\mu\nu}(\mbox{zero mode})\approx {1\over z^2}.  
\end{equation}
An invariant obtained by contracting the 
Weyl tensor with itself $C_{\mu\nu\rho\sigma}C^{\mu\nu\rho\sigma}$ 
behaves as $\approx z^4$. 
Thus, the infinity in the direction of the extra dimension 
is a curvature singularity. 
For the massive modes the situation is worse. 
The wave function behaves as 
\begin{equation}
 h_{\mu\nu}(\mbox{massive mode})\approx {1\over \sqrt{z}}.  
\end{equation}
Hence, the same invariant more severely diverges as 
$C_{\mu\nu\rho\sigma}C^{\mu\nu\rho\sigma}\approx z^7$.

However, such a divergence does not indicate
the breakdown of the perturbation analysis. 
The perturbed metric induced by the matter 
fields on the brane consists of a superposition of various modes. 
For the static case, the 5D Green function is 
approximately evaluated in Ref.\cite{GarTan}. 
The result clearly showed that perturbations at large $z$ 
are regular. Also in the dynamical cases, the asymptotic regularity 
of perturbations was shown in Ref.\cite{Tanaka:2000zv}.
An interesting point which we wish to stress here 
is that perturbations become 
regular only after summation over all massless and massive modes. 

Non-linear perturbations in this model are more complicated. 
If we adopt the picture of the 4D effective theory with 
a tower of massive gravitons, one may think that 
the higher order perturbations can be treated by  
taking into account the effective coupling between 
gravitons with various masses. However, this approach 
does not work. Let us consider the 
three point interaction vertex. The effective 
coupling constant between various massive 
gravitons will be obtained by 
expanding the action ($\propto {}^{(5)}R$) to 
the third order with respect to the 
metric perturbation $h_{ab}$ and integrating out 
the dependence on the extra dimension. 
The quantity to be calculated will take 
the form 
\begin{equation}
 \int d^4 x\int_{\ell}^{\infty} dz\,
   \sqrt{-g} g^{**} g^{**} g^{**} g^{**} 
          h_{**,*} h_{**,*} h_{**}. 
\label{vertex}
\end{equation}
The asymptotic behavior of respective component is given by 
$\sqrt{-g} \sim z^{-5}$ and $g^{**} \sim z^2$. If we substitute 
the massive mode wave function, $h_{**}$ 
and also $h_{**,*}$ are $\sim z^{-1/2}$. 
Hence, the integrand of (\ref{vertex}) behaves as $\propto z^{3/2}$, 
and the $z$-integration does not converge. 
Therefore we cannot define the effective coupling constant 
in this manner. 

Nevertheless, this does not directly imply any 
catastrophe at least at the classical level. 
For the static and spherically symmetric 
configurations in the 4D sense, 
second order perturbations were calculated to show that 
the perturbations behave well, and the correction to 4D general relativity 
is suppressed by the factor of $O(\ell^2/r_{\star}^2)$\cite{GiaRen,KudTan}. 
The approximate reproduction of the results for 4D 
general relativity is also confirmed numerically in
Ref.\cite{Wiseman}, in which the strong gravity regime was 
also investigated. 
Hence, one may be able to conclude that the gravity in the 
RS infinite braneworld is well 
approximated by 4D general relativity, 
although non-linear perturbations 
have not been computed in dynamical cases at all\cite{Roy}.  

\section{Black hole and AdS/CFT correspondence}

In the preceding section we have observed that the
induced metric on the RS brane mimics 
the results of 4D general relativity well. This seems to 
work even in the strong gravity regime. 
However, no black hole solution which is asymptotically 
AdS has been found so far. 

In Ref.\cite{ChaHaw}, a black string solution given by 
\begin{equation}
 ds^2={\ell^2\over (|z|+\ell)^2}\left[dz^2+q_{\mu\nu}^{(4)}dx^{\mu} dx^{\nu}
 \right], 
\end{equation}
was discussed. Here $q_{\mu\nu}^{(4)}dx^{\mu} dx^{\nu}$ is the 
usual 4D Schwarzshild metric.
The induced geometry on the brane at $z=0$ 
is exactly 4D Schwarzshild spacetime. However, 
the asymptotic 
value of $C_{\mu\nu\rho\sigma}C^{\mu\nu\rho\sigma}$ behaves
$\propto z^4 r^{-6}$, where 
$r$ is the Schwarzshild radial coordinate. If we take the 
$z\to \infty$ limit for a fixed $r$, this curvature invariant 
diverges\cite{ChaHaw}.
Also, this configuration is unstable\cite{GreLaf,Gregory,HorMae}. 
Hence, the black string solution will not be 
an appropriate candidate for the final state of the gravitational 
collapse in the RS braneworld. A conjecture raised in Ref.~\cite{ChaHaw} 
is that there will be a configuration called ``black cigar'' 
for which there is an event horizon localized near the brane. 

However, as we have mentioned above, 
no black hole solution which is asymptotically 
AdS has not been found so far, although there were 
several works aiming at finding it. 
Here we suggest that the brane black hole may not exist, 
based on the argument of AdS/CFT correspondence.  
For an introduction to AdS/CFT correspondence, 
we follow Refs.\cite{HawHer,ShiIda}. 
Here we consider the RS model 
without matter fields on the brane.  
The AdS/CFT correspondence implies the relation 
\begin{equation}
 S_{RS}=S_{EH}^{(4)}+2W_{CFT}, 
\end{equation}
where $W_{CFT}$ is the connected 
Green function with a high frequency cutoff 
for certain 4D CFT fields evaluated on the background 
metric induced on the brane, and  
\begin{eqnarray}
S_{EH}^{(4)} & = & -{\ell \over 16\pi G_5}
              \int d^4 x\sqrt{-q}^{(4)}R,  
\end{eqnarray}
is the ordinary 4D Einstein-Hilbert action for the induced metric 
on the brane. 
This formula indicates that the RS infinite 
braneworld is equivalently described by 4D general relativity  
coupled to conformal fields. The number of degrees of freedom of 
the conformal fields is 
$O(\ell^2/G_4)$, which is supposed to be large. 

It is known that the quantum effect
by means of 4D CFT corresponds to the classical effect due to 
the bulk graviton in 5D picture. 
This fact can be also understood in 
the following way. Let us consider the energy momentum tensor 
of the 4D CFT due to the vacuum polarization effect induced by 
the curved geometry. 
In this case, the contribution to the 
energy momentum tensor from each field will be  
$O(1/L^4)$, where $L$ is the characteristic length scale 
of the spacetime curvature. 
Thus, the vacuum polarization part of the 
energy momentum tensor $^{(4)}T^{(Q)}_{\mu\nu}$ in total 
will become $O(\ell^2/G_4 L^4)$, where we have taken into 
account the number of degrees of freedom.  
Hence, from 4D Einstein equations 
$^{(4)}\Box h_{\mu\nu}\approx G_{4} {}^{(4)}T_{\mu\nu}$, 
the additional metric perturbation caused by this 
energy momentum tensor $^{(4)}T^{(Q)}_{\mu\nu}$ 
is estimated as $h_{\mu\nu}^{(Q)}=O(\ell^2/L^2)$. 
On the other hand, 
the effective energy momentum tensor induced by 
the quantum effect in the 5D point of view 
is also given by the curvature scale of 
the spacetime. When we discuss in 5D picture, 
there are two characteristic length scales, $\ell$ and $L$. 
We denote both of them by $\tilde L$ without distinguishing them.  
Then, we will have 
$^{(5)}T_{\mu\nu}^{(Q)}\approx \tilde L^{-5}$. 
Then, from 5D Einstein equations,  
$^{(5)}\Box h_{\mu\nu}\approx G_{5} {}^{(5)}T_{\mu\nu}$, 
we will obtain 
${h}_{\mu\nu}^{(Q)}=O(\tilde L^{-3} G_5)=O(\tilde L^{-2} G_4)$. 
Note that the number of degrees of freedom is a few 
in this case. 
Since the power of $G_{4}$ does not coincide 
in the above two expressions for $h_{\mu\nu}^{(Q)}$, 
it is almost impossible to expect that the contribution due to 
the quantum effect in 5D picture corresponds to that in 4D CFT picture. 
This mismatch comes from the large number of 
CFT fields of $O(\ell^2/G_4)$. 

The correspondence at the level of  
linear perturbation was made explicit in Ref.\cite{DufLiu}. 
The leading correction to 4D general relativity starts with  
$O(\ell^2/L^2)$ in $h_{\mu\nu}$, and this leading term was shown to 
satisfy the correspondence relation. The possible error of 
the correspondence is higher order in $\ell$. 

It is very difficult to verify the correspondence beyond the linear 
perturbation. However, there is a supporting evidence 
related to the trace anomaly. Here we follow Ref.\cite{ShiIda}.  
The trace of the effective Einstein equations (\ref{4DEeq}) becomes 
\begin{equation}
^{(4)}G=8\pi G_4 T+
{(8\pi G_4 \ell)^2\over 4}\left(T_{\mu\nu}T^{\mu\nu}-{1\over 3}T^2\right),
\label{trace1}
\end{equation}
where the second term on the right hand side comes from the trace of
$\pi_{\mu\nu}$. 
On the other hand, 
the trace part of the energy momentum tensor of CFT
is solely determined by the trace anomaly, and hence the 
trace of the Einstein equation with CFT becomes 
\begin{equation}
^{(4)}G=8\pi G_4 T+
  {\ell^2/4} 
\left({}^{(4)}G_{\mu\nu}{}^{(4)}G^{\mu\nu}
     -{1\over 3}{}^{(4)}G\,{}^{(4)}G\right).
\label{trace2}
\end{equation}
In the situation that the correction to 4D general relativity is 
suppressed by the factor of $O(\ell^2/L^2)$, we have 
${}^{(4)}G_{\mu\nu}=8\pi G_4 T_{\mu\nu}+O(\ell^2/L^4)$. 
Then the difference between (\ref{trace1}) and (\ref{trace2}) becomes 
$O(\ell^4/L^6)$. Hence, the AdS/CFT correspondence for the 
trace part of the Einstein equation holds for the leading correction 
proportional to $\ell^2$ even in the non-linear perturbation.  

Now, let us apply the argument of the AdS/CFT correspondence 
to the formation of a black hole in the RS 
braneworld. 
In 4D CFT picture, a black hole is formed under the 
presence of a large number of conformal fields.  
Then, the back reaction due to the Hawking radiation 
will be much more efficient than in the ordinary 4D theory by 
the factor of $\ell^2/G_4$. 
If the AdS/CFT correspondence is valid in this situation, 
the quantum back reaction due to the Hawking radiation in 4D picture 
must be described as a classical dynamics in 5D picture. 
This means that the black hole evaporates as a classical process
in 5D picture.
This may implies that there is no stationary black hole 
solution in the 5D RS model. 
The result obtained in Ref.\cite{Roy} that the exterior of the
dynamically collapsing star cannot be static is also in line with 
the present conjecture.  
Of course, when the size of black hole 
becomes as small as the AdS curvature radius $\ell$, the correspondence 
may cease to hold. 

There is a static black hole solution 
when a 2-brane in 4 dimensional bulk is considered. 
One may think this lower dimensional example 
as an evidence against the conjectured absence of black hole
solution in braneworld.  
This solution was found by 
Emparan, Horowitz and Myers\cite{Emparan}. 
The 3-dimensional metric induced on the brane looks 
similar to the 4-dimensional Schwarzshild black hole;
\begin{equation}
 ds^2=-\left(1-{2\mu\ell\over r}\right) dt^2 
      +\left(1-{2\mu\ell\over r}\right)^{-1} dr^2 
      +r^2 d\varphi^2. 
\end{equation}
The period of identification in  
$\varphi$-direction is not $2\pi$ 
but $\Delta\varphi\approx \displaystyle{
4\pi\over 3(2\mu)^{1/3}}$, where we assumed that 
$\mu\gg 1$. 
For this black hole geometry, we have 
\begin{equation}
 E_{\mu\nu}={\mu\ell\over r^3}\, \mbox{diag}(1,1,-2). 
\label{3d}
\end{equation}
If we apply the AdS/CFT correspondence,  
the energy momentum tensor of CFT is estimated by  
$T_{\mu\nu}^{CFT}\approx -(8\pi G_3)^{-1} E_{\mu\nu}$. 
To maintain a static configuration 
under the existence of the Hawking radiation,  
there should exist thermal bath which supplies the incoming 
energy flux to balance with the Hawking radiation. 
However, $T_{\mu\nu}^{CFT}$ estimated above decays too fast for large $r$ 
to explain this thermal bath. 

However, 
we do not think that this is 
a counter example against the conjecture of the classical evaporation 
of black holes in the RS braneworld. 
An important difference of this lower dimensional example 
from the 4D Schwarzshild black hole 
is that the induced metric is not similar to a solution of 3D vacuum 
Einstein equation. 
It will be also worth pointing out that 
this effective energy momentum tensor (\ref{3d}) 
can be understood as the Casimir energy of CFT.  
This spacetime is very compact in the $\varphi$-direction. The period 
in this direction is 
$r\Delta\varphi\approx r/\mu^{1/3}$, and is much shorter than 
the curvature scale $L\approx (\mu\ell/r^3)^{-1/2}$ even at the horizon. 
Hence, the approximation by a cylinder 
with a fixed radius $\approx r/\mu^{1/3}$ will work. 
For such a configuration, 
the energy momentum tensor for one conformal field is given by 
$
 \approx (\mu/r^3)\mbox{diag}(1,1,-2). 
$
The number of CFT species of $O(\ell/G_3)$ consistently 
explains the correspondence relation.

\section{summary and discussion}

In this paper, we reviewed the studies on the gravity in the 
Randall-Sundrum infinite braneworld. 
All the computations performed so far suggest that  
this model recovers the 4D general relativity as an 
effective theory induced on the brane. 
However, no black hole solution with the regular 
asymptotic behavior has been obtained.  
We discussed the possibility that there is no static 
black hole solution in this model,  
applying the argument of the AdS/CFT correspondence. 

The first question is whether a brane black hole looks like an ordinary 
4D BH or not. If possible black hole solutions are quite different from the 
ordinary black hole in 4D general relativity, the situation
will be very interesting because we may have a chance to
probe the extra-dimension by astronomical observations of black holes. 
If the solutions are similar to the ordinary ones, the next question
arises whether the AdS/CFT Correspondence applies for the 
black hole configuration. 
The case that the correspondence does not hold 
in such a strong gravity regime is still interesting 
because it may be possible to distinguish 
the model described by 4D CFT picture 
from the 5D RS model observationally in this case. 
The remaining 
possibility that the answer to the above question is "YES", i.e., 
that the correspondence applies even for a black hole configuration, 
is extremely interesting.  

In this case, we can evaluate
the mass loss rate due to the Hawking radiation as 
$\dot M/M\sim N\times (1/G_4^2 M^3)\sim \ell^2/(G_4 M)^3$. 
Then the evaporation time scale becomes 
$\tau=(M/M_\odot)^3(1mm/\ell)^2\times 1$ year. 
The existing black holes in X-ray binaries will give 
a stronger constraint on the value of $\ell$ 
than that from the laboratory experiment.  
The energy lost, say, for the last 10 second of the evaporation 
will become
$E=(\ell/1mm)^{2/3} \times 10^{52}$ erg. 
The amount of energy can be very large.  
However, this energy is radiated into CFT (= KK gravitons), 
which are almost decoupled from the ordinary matter fields. 
Hence, the direct observation of this energy seems to be difficult. 

What does this evaporation look like in 5d picture? 
Although this is just a speculation, we would like to 
propose one possible explanation. 
We assume that blobs of the size $\ell^3$ are continuously 
created at the dynamical time scale of the BH, $G_4 M$, with the 
area of 5D horizon kept constant. The size of the blobs is 
suggested by the maximum length scale for the 
instability of black string\cite{GreLaf,Gregory}. 
Since the area of the main part of the horizon close to the brane 
will be $\approx M^2\ell$, 
we have ${d\over dt} M^2\ell\approx \ell^3/G_4 M$.
From this estimate, we obtain the same order of magnitude 
for the mass loss rate as before. 
If the blobs escape to the 5th direction, this picture 
will consistently explain the propagation of CFT energy to the 
spatial infinity in 4D picture\cite{RubGre}. 

This correspondence, if it holds, can have an important meaning 
even in the case that our real universe is not described by the RS braneworld. 
Since 4D black hole evolution including back reaction 
due to the Hawking radiation can be described by a 
5D classical problem in this case, 
it may become possible to 
solve the back reaction problem by using 5D numerical relativity. 
If we assume spherical symmetry in 4 dimensional sense, we have only 
to solve a problem with axial symmetry in 5D picture.

\vspace{5mm}
\centerline{\bf Acknowledgements}
\vspace{5mm}
The author would like to thank J. Garriga for useful comments
and discussions. To complete this work, the discussion during and 
after the YITP workshop YIYP-W-01-15 on ``Braneworld - Dynamics of 
spacetime boundary'' was useful. In particular, the author thanks 
R. Maartens and N. Kaloper for their useful comments. 
This work was supported in part by the Monbukagakusho Grant-in-Aid
No.~1270154 
and by the Yamada Science Foundation.


\begin{thebibliography}{99}
\bibitem{Horava:1996qa}
P.~Horava and E.~Witten,
Nucl.\ Phys.\ B {\bf 460}, 506 (1996);
\\
P.~Horava and E.~Witten,
Nucl.\ Phys.\ B {\bf 475}, 94 (1996).


\bibitem{Arkani-Hamed}
N.~Arkani-Hamed, S.~Dimopoulos and G.~Dvali,
Phys.\ Lett.\ B {\bf 429}, 263 (1998);
\\
I.~Antoniadis, N.~Arkani-Hamed, S.~Dimopoulos and G.~Dvali,
Phys.\ Lett.\ B {\bf 436}, 257 (1998).

\newpage


\bibitem{Randall:1999ee}
L.~Randall and R.~Sundrum,
Phys.\ Rev.\ Lett.\ {\bf 83}, 3370 (1999).


\bibitem{Randall:1999vf}
L.~Randall and R.~Sundrum,
Phys.\ Rev.\ Lett.\ {\bf 83}, 4690 (1999).

\bibitem{Shiromizu:2000wj}
T.~Shiromizu, K.~Maeda and M.~Sasaki,
Phys.\ Rev.\ D {\bf 62}, 024012 (2000).

\bibitem{GarTan}
J.~Garriga and T.~Tanaka,
Phys.\ Rev.\ Lett.\ {\bf 84}, 2778 (2000).

\bibitem{Giddings:2000}
S.B.~Giddings, E.~Katz and L.~Randall,
JHEP {\bf 0003}, 023 (2000).


\bibitem{Tanaka:2000zv}
T.~Tanaka,
Prog.\ Theor.\ Phys.\  {\bf 104}, 545 (2000).

\bibitem{GiaRen}
I.~Giannakis and H.~c.~Ren,
Phys.\ Rev.\ D {\bf 63}, 024001 (2001).

\bibitem{KudTan}
H.~Kudoh and T.~Tanaka,
Phys.\ Rev.\ D {\bf 64}, 084022 (2001).

\bibitem{Wiseman}
T.~Wiseman,
hep-th/0111057.
 
\bibitem{Roy}
A limitted but interesting analysis on gravitational collapse 
on the brane was done by \\
M.~Bruni, C.~Germani and R.~Maartens,
Phys.\ Rev.\ Lett.\  {\bf 87}, 231302 (2001).

\bibitem{ChaHaw}
A.~Chamblin, S.~W.~Hawking and H.~S.~Reall,
Phys.\ Rev.\ D {\bf 61}, 065007 (2000).

\bibitem{GreLaf}
R.~Gregory and R.~Laflamme,
Phys.\ Rev.\ Lett.\ {\bf 70} 2837 (1993). 

\bibitem{Gregory}
R.~Gregory,
Class.\ Quant.\ Grav.\ {\bf 17}, L125 (2000).

\bibitem{HorMae}
G.~T.~Horowitz and K.~Maeda,
Phys.\ Rev.\ Lett.\  {\bf 87}, 131301 (2001).

\bibitem{HawHer}
S.W.~Hawking, T.~Hertog and H.S.~Reall
Phys.\ Rev.\ D {\bf 62} 043501 (2000). 

\bibitem{ShiIda}
T.~Shiromizu and Daisuke Ida
Phys.\ Rev.\ D {\bf 64} 044015 (2001). 

\bibitem{DufLiu}
M.~J.~Duff and J.~T.~Liu,
Phys.\ Rev.\ Lett.\  {\bf 85}, 2052 (2000).

\bibitem{Emparan}
R.~Emparan, G.~T.~Horowitz and R.~C.~Myers,
JHEP{\bf 0001}, 007 (2000).

\bibitem{RubGre}
R.~Gregory, V.A.~Rubakov and S.M.~Sibiryakov
Class.\ Quant.\ Grav.\ {\bf 17} 4437-4450 (2000). 

\end{thebibliography}
\end{document}